\documentclass[submission,copyright,creativecommons]{eptcs}
\providecommand{\event}{DICE-FOPARA 2019} % Name of the event you are submitting to
\usepackage{breakurl}             % Not needed if you use pdflatex only.
\usepackage{underscore}           % Only needed if you use pdflatex.

\usepackage[numbers,sort&compress]{natbib} % sort multi-citations by number - [3, 5] instead of [5, 3]
\usepackage{tikz}

\usetikzlibrary{arrows.meta}
\usetikzlibrary{fit}
\usetikzlibrary{positioning}

\usepackage{analyse}

\title{Type-Based Resource Analysis on Haskell}
\author{Franz Siglm\"uller
\institute{Ludwig Maximilian University\\ Munich, Germany}
\email{frsiglmu@web.de}
}
\def\titlerunning{Type-Based Resource Analysis on Haskell}
\def\authorrunning{Franz Siglm\"uller}

\begin{document}
\maketitle

\begin{abstract}
We propose an amortized analysis that approximates the resource usage of a Haskell expression.
Using the plugin API of GHC, we convert the Haskell code into a simplified representation called GHC Core.
We then apply a type-based system which derives linear upper bounds on the resource usage.
This setup allows us to analyze actual Haskell code, whereas previous implementations of similar analyses do not support any commonly used lazy functional programming languages.
%For this purpose, we implemented a type-based system which 
%This is implemented as a type-based system which 
%Using the plugin API of the Haskell compiler GHC and an LP solver, we can derive annotated types for a given Haskell expression.
%These type annotations indicate upper bounds on the resource usage of the program.
%This allows us to analyze real Haskell code;
%Unlike previous work, which does not support any commonly used lazy functional programming languages.
\end{abstract}

\section{Introduction}
\label{introduction}

It can be difficult to estimate the runtime costs of a program written in a lazy programming language.
For example, consider the following two versions of a ``repeat'' function, which returns an infinite list with a given element:

\smallskip

\begin{minipage}{\textwidth}
	
	\hspace{2em}\verb$repeat x = let xs = x : xs in xs$
	
	\hspace{2em}\verb$it  = repeat  1 :: [Int]$
	
	\hspace{2em}\verb$repeat' x = x : repeat' x$
	
	\hspace{2em}\verb$it' = repeat' 1 :: [Int]$
	
\end{minipage}

\smallskip

To a novice, it may not be obvious that the memory usage of this function is constant for the first version and linear for the second version.
To assist in detecting these differences, we want to automatically generate annotated types such as the following:

\smallskip

\begin{minipage}{\textwidth}
	
	\hspace{2em}$\rtyps40 \mathit{it} : \mu X.~ \{\lkey{[]} : (0, []) ~|~ \lkey{(:)} : (0, [\thunkq0{\lkey{Int}}, \thunkq0X])\}$
	
	\hspace{2em}$\rtyps60 \mathit{it'} : \mu X.~ \{\lkey{[]} : (0, []) ~|~ \lkey{(:)} : (0, [\thunkq0{\lkey{Int}}, \thunkq2X])\}$
	
\end{minipage}

\smallskip

In these types, the second field of the $\lkey{(:)}$ constructor is wrapped in a thunk type that is annotated with costs 0 and 2, respectively.
These values represent additional costs -- in this case, memory allocations -- that arise from accessing subsequent nodes of the linked list.
As this cost is 0 in the type of the first version, this means that no additional allocations will take place and the list therefore requires constant space.
For the second version, the value is greater than 0, indicating linear costs.

In this paper, we describe our implementation of a tool that can derive these annotated types automatically.
Given a Haskell module, our tool will use the GHC compiler to parse the code and translate it into an intermediate representation which is more easy to analyze.
We then apply our own custom type-based system on each expression to derive an annotated typing which depicts an upper bound on the number of resources required for evaluation.
Depending on the cost model used, this can be used to approximate memory usage or runtime duration, for example.

%TODO: The structure of the remaining paper is as follows ...

This paper is a summary of the master's thesis ``Implementation of an Automated Amortized Analysis on GHC Core as Compiler Plugin''. \cite{ImplementationOfAnAutomatedAmortizedAnalysisOnGHCCoreAsCompilerPlugin}
Note that sections of this paper may be similar or identical to that thesis.

\section{Related Work}

Our work is rooted in Steffen Jost's PhD thesis ``Automated Amortized Analysis'' \cite{AutomatedAmortisedAnalysis}.
In his work, Jost introduced a custom language Arthur which can be analyzed statically using a custom type system.
For any given Arthur expression, this system can derive a type annotated with numeric variables, as well as a linear program over these variables.
After feeding this linear program to an LP solver, an annotated type can be generated which represents linear upper bounds on the memory usage.

A variation of this system was later integrated into the OCaml compiler by Jan Hoffmann et al. \cite{ResourceAwareML, TowardsAutomaticResourceBoundAnalysisForOCaml}
This allows them to analyze real OCaml code by using the existing compilation pipeline to parse the original code, and then applying their own analysis on the intermediate code generated by the compiler.

However, these previous systems are designed around strict programming languages and cannot be applied to lazy languages such as Haskell in a meaningful way.
While delayed evaluation is supported to some extent -- for example, using function abstraction and application --, this needs to be encoded explicitly within the analyzed expression.
In lazy languages such as Haskell, however, it is implicit that we need to treat every sub-expression as lazy.
As an example of this distinction, consider the \texttt{repeat'} function provided in the previous section:
This definition is only meaningful in a lazy language, where only the required prefix of the infinite list will be evaluated;
While a strict interpretation of the same implementation will attempt to evaluate the full infinite list at once, triggering an infinite loop.

A new type-based system designed around lazy evaluation was proposed in ``Type-based Cost Analysis for Lazy Functional Languages'' by Steffen Jost, Pedro Vasconcelos, M{\'a}rio Florido and Kevin Hammond. \cite{TypeBasedCostAnalysisForLazyFunctionalLanguages}
This is achieved by introducing ``thunk types'' which are used for sub-expressions that may be evaluated lazily.
Thunk types are annotated with the costs required for evaluating the corresponding sub-expression;
This cost has to be payed only when the expression is actually evaluated, and can be discarded otherwise.
This system has later been examined and extended as part of several bachelors' theses. \cite{EvaluationUndErweiterungEinerTypBasiertenKostenanalyseFuerFunktionaleSprachenMitVerzoegerterAuswertung, ErweiterungDesPrototypsEinerTypBasiertenKostenanalyseFuerFunktionaleSprachenMitVerzoegerterAuswertung, MutualRecursiveDefinitionsForATypeBasedCostAnalysisForLazyFunctionalLanguages}
Most notably, these extended the language and the type system to introduce support for mutual recursion.

The major drawback of this system is the programming language which is analyzed.
Just like Arthur before, Jost et al.\ introduced their own (unnamed) custom programming language which we have dubbed ``JVFH''.
This language is unwieldy to use, as the primary goal in designing this language was being convenient to analyze rather than to write programs with.
For example, in order to analyze the \texttt{repeat'} function given before, we first have to translate it into JVFH as follows:

\smallskip

\begin{minipage}{\textwidth}
	
	\hspace{2em}\verb$let repeat' = \x -> let xs = repeat' x$
	
	\hspace{2em}\verb$                    in letcons xt = Cons (x,xs)$
	
	\hspace{2em}\verb$                       in xt$
	
	\hspace{2em}\verb$in let one = 1$
	
	\hspace{2em}\verb$   in let it' = repeat' one$
	
	\hspace{2em}\verb$      in it'$
	
\end{minipage}

\smallskip

While it is possible to use the JVFH analysis for Haskell in this way, performing this translation step manually is time-consuming and error-prone.
In our work, we will circumvent this issue by adapting the existing JVFH type system to one of the intermediate languages used within the GHC compiler.
This means that the translation can be performed automatically by using the existing GHC compilation pipeline.

\section{GHC Architecture}
\label{ghcarchitecture}

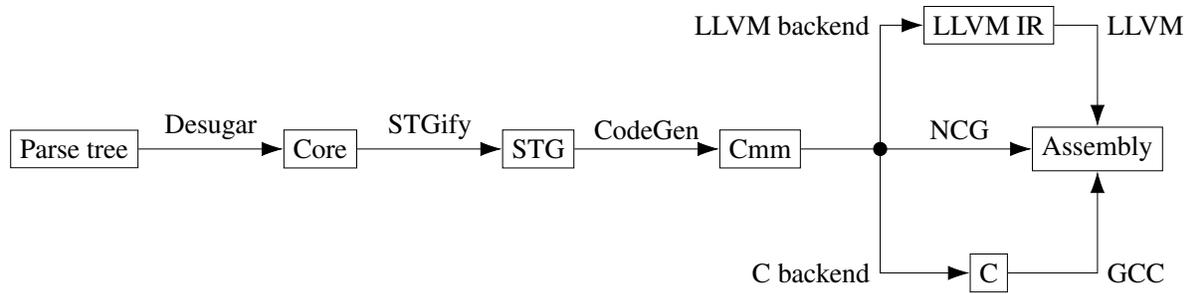
\begin{figure}
	\centering\resizebox{\linewidth}{!}{
		\begin{tikzpicture}
	\begin{scope}[every node/.style={rectangle,draw=black}]
	    \node (Parse) at (0,0) {Parse tree};
	    \node[right = 2cm of Parse] (Core) {Core};
	    \node[right = 2cm of Core] (STG) {STG};
	    \node[right = 2cm of STG] (CMM) {Cmm};
	    \node[right = 1cm of CMM, shape = circle, fill=black, scale=0.5] (Fork) {};
 	    \node[right = 2cm of Fork] (Assembly) {Assembly};
 	    \node(WrapNCG) [rectangle, draw=none, fit = (Fork.north) (Fork.south) (Assembly.north) (Assembly.south)] {};
	    \node[above = 1cm of WrapNCG] (LLVM) {LLVM IR};
	    \node[below = 1cm of WrapNCG] (C) {C};
	\end{scope}
	
	\draw (CMM) -- (Fork);
	
	\begin{scope}[-{Latex[length=3mm,width=2mm]}]
		\draw (Parse) -> (Core) node [midway, above] {Desugar};
		\draw (Core) -> (STG) node [midway, above] {STGify};
		\draw (STG) -> (CMM) node [midway, above] {CodeGen};
		\draw (Fork) |- (LLVM) node [midway, left] {LLVM backend};
		\draw (LLVM) -| (Assembly) node [midway, right] {LLVM};
		\draw (Fork) |- (C) node [midway, left] {C backend};
		\draw (C) -| (Assembly) node [midway, right] {GCC};
		\draw (Fork) -> (Assembly) node [midway, above] {NCG};
	\end{scope}
\end{tikzpicture}
	}
	\caption{Basic overview of the GHC compilation pipeline. Based on a graphic in \cite{IKnowKungFuLearningStgByExample}.}
	\label{compilationpipeline}
\end{figure}

In this section, we provide a basic summary of the compilation pipeline used within GHC, depicted in Figure \ref{compilationpipeline}.
The original Haskell code is first parsed into an abstract syntax tree, which is then type-checked.
It is then desugared and translated to GHC Core, a lazy functional language based on System F.
At this point, the compiler will apply most of its code optimizations.
Afterwards, the code is translated to STG, a subset of GHC Core minus certain type information which is not needed anymore after this point.
It is then translated into a low-level imperative language called Cmm, before one of several different backends can be used to translate it into Assembler and, finally, the finished binary.

For our analysis, we will be looking at GHC Core, as it is the most similar to the JVFH language.
Additionally, the other intermediate languages used by GHC Core have a number of drawbacks that make them unsuitable for our analysis:

The syntax of the parse tree is a considerably more redundant than that of GHC Core;
In fact, the type used in GHC to represent a Core expression consists of only 10 different constructors \cite{TheGhcApiCoreSyn}, while the Haskell syntax is represented by around 50 different constructors \cite{TheGhcApiHsExpr}.
While it is possible to implement a type system on the parse tree, it is tedious and does not provide any advantage for our purpose.
Additionally, most of the compiler optimizations are applied only after the translation to GHC Core.
It is preferable to analyze this optimized code, as it is more representative of the final compiled binary.

The STG code is unsuitable for our purpose, because it lacks certain type information such as type abstraction and application.
As GHC Core is an extension to System F and STG is a subset of Core missing type information, we assume that the undecidability of type inference in System F \cite{TypabilityAndTypeCheckingInSystemFAreEquivalentAndUndecidable} also applies to STG;
This is not an issue for the compiler, as type checking has already completed at this point and the compiler does not need the discarded type information anymore.
However, as our analysis is based on a type system, this information is essential to us in fully supporting the language, particularly polymorphism. 
While our current implementation does not support polymorphism yet and therefore may also work on STG, choosing STG over Core would severely hamper any future work on supporting the full language.

Finally, Cmm and any of the languages used thereafter are strict imperative languages.
Thus, they are unsuitable for an adaption of the JVFH type system, which is designed around a lazy functional language.
Analyzing these languages may be feasible, but this would require a completely different approach.

The GHC compiler has an API for custom plugins that can be used to interact with the compilation pipeline.
Plugins can be implemented by writing a Haskell module that exports variables with specific names and specific types, and they can be used by calling GHC with specific command line arguments. \cite{GhcUsersGuideDocumentation822, TheGhcApiPlugins}

When we started our implementation, three different types of plugins where supported:
Core plugins can modify the optimization pipeline applied to the Core intermediate code;
Frontend plugins can take full control over the entire compilation pipeline;
And typechecking plugins can enhance the capabilities of the constraint solver used during type checking.
The latter has proven unsuitable for our purpose, as we need to establish our own type system, not enhance the existing one.
We have implemented both a Core and a frontend plugin which serve to obtain a copy of the Core intermediate code and pass it to our analysis.
This future-proofs our tool in case one of the APIs is removed in the future.

\section{Language \& Type Syntax}

\begin{figure}
	\input{gc-language}
	\caption{Simplified representation of the GHC Core syntax}
	\label{gc-language}
\end{figure}

In the next section, we will discuss how we had to adapt the existing JVFH type rules for GHC Core.
For this purpose, we will be using a streamlined representation of the Core syntax as depicted in Figure \ref{gc-language}.
This representation is slightly different from the one actually used within the compiler \cite{TheGhcApiCoreSyn}.
For example, we distinguish more clearly between variables and constructors, both of which are represented as \texttt{Var} within the GHC compiler.
On the other hand, we also omit several expressions which are currently unsupported in our analysis, namely \texttt{Cast}, \texttt{Coercion} and \texttt{Tick}.

\begin{figure}
	\input{prev-types}
	\caption{Type syntax, as originally defined in \cite{TypeBasedCostAnalysisForLazyFunctionalLanguages}}
	\label{prev-types}
\end{figure}

For our types, we are reusing the type syntax from the original JVFH system without modifications.
This syntax, as depicted in Figure \ref{prev-types}, covers type variables, functions, thunks and algebraic types.
Most of these contain annotations that signify the resource usage of the respective expression.
Depending on the cost model used, this can refer to memory usage or runtime duration, for example.

The type $\typAr AqB$ represents a function, that takes an argument of type $A$, and returns a result of type $B$.
The type annotation $q$ signifies the cost of executing this function, i.e., evaluating the function result to weak head normal form.

The algebraic type \mbox{$\mu X.\; \{\, c_1:(q_1,\vec{A}_1) \;|\; \cdots \;|\; c_n:(q_n,\vec{A}_n) \,\}$} is a possibly recursive type consisting of several constructors $c_i$.
For each constructor, we specify a potential $q_i$, which are resources ``reserved'' during structure allocation for future use.
During a pattern match, this potential can be redeemed to pay for any upcoming calculations.
Additionally, each constructor is also associated with a list of constructor fields $\vec{A}_i$.
The type variable $X$ can be used within these constructor fields as a recursive reference to the algebraic type.
As an example, consider the following representation of a list type:

$\mu X.\; \{\lkey{Nil}: (1, []) \;|\; \lkey{Cons}: (2, [\thunkq3{\lkey{Int}}, \thunkq4X])\}$

This example type consists of two constructors:
The first, $\lkey{Nil}$, has potential 1 and no fields;
and the second, $\lkey{Cons}$, has potential 2 and two fields, namely a list element of type $\lkey{Int}$, and a reference to the next list node, represented recursively with a type variable $X$.
Note that type variables in our system may only be used as recursive references, and therefore must always be contained within algebraic types that bind them.

Finally, the thunk type $\thunkq qA$ is a wrapper that represents lazy evaluation;
It signifies that an expression of this type possibly has not been evaluated to weak head normal form yet, and doing so will use up $q$ resources.
Thunk types can only (and will always) appear in specific places:
Namely, as the argument of any function type, as the constructor fields of any algebraic type, and as the type of every variable in the context of any typing judgment.

\begin{sloppypar}
We apply some workarounds for Haskell/Core types that do not have any direct equivalent in this syntax:
Primitive types such as ``\texttt{Int\#}'' or ``\texttt{Float\#}'' are represented as an empty algebraic type \mbox{``$\mu X.\{\}$''} in our system;
and type abstractions ``\texttt{forall x.\ T}'' are replaced with an ``artificial'' function ``$\typAr{\mu X.\{\}}0T$''.
However, note that polymorphism is currently unsupported in our implementation;
Therefore, this artificial function does not actually have any meaningful effect within the type system, and only serves as an informatory annotation for the user.
Also note that this representation breaks the requirement that all function arguments are wrapped in thunk types;
This can be used to easily distinguish ``real'' functions from an artificial function representing type abstraction.
\end{sloppypar}

Applying these workarounds instead of extending the type syntax may seem restricting, but it offers one major advantage:
The JVFH type system defines several relations between different types, which we can reuse without any modifications.
This would not be possible if we modified the type syntax or its semantics.

\section{Type Rules}
\label{typerules}

In this section, we will give an overview over the adapted type rules used in our implementation.
However, note that we abstain from reciting a large portion of the rules and definitions which we reuse from the original JVFH system without modification.
Instead, please refer to the original publication by Jost et al. \cite{TypeBasedCostAnalysisForLazyFunctionalLanguages}

The following types of judgments are used in both the JVFH and our type system:

\begin{itemize}
\item
	The typing judgment $\Gamma \rtyps pq e : T$ states that, given context $\Gamma$, expression $e$ has type $T$.
	Additionally, $p$ is an upper bound on the number of resources required to evaluate $e$ to weak head normal form;
	and $q$ is a lower bound on the number of resources that are available after this evaluation.
	
	Among the type rules that can be applied to these judgments, we distinguish between syntax driven rules and structural rules.
	The latter serve various difference purposes;
	For example, the \rnty{Prepay} rule can be used to pay the thunk cost of any variable only once, even if the variable is used multiple times in the expression;
	And the \rnty{Relax} rule allows us to increase the over-approximation of the cost upper bounds in a typing judgment on purpose, thereby relaxing linear constraints which otherwise might render the linear program unsolvable.
	While the purpose of the various structural type rules vary widely, they have in common that they can be applied to any typing judgment, regardless of the expression.
	This means that we can reuse these rules in our system without any modifications.
	
	The syntax driven rules can only be applied to typing judgments with specific expressions.
	Many of these rules contain cost constants, which can be set by the user;
	The basic goal of the type system is to count how often certain kinds of expressions are executed, and to add up the cost constants associated with these expressions.
	For example, if we set the cost constants associated with any expressions involving variable bindings to 1, and the remaining constants to 0, our analysis will derive an upper bound on the number of memory allocations required to evaluate an expression.
	However, as these type rules directly depend on the available expressions of the language and their syntax, the original JVFH rules cannot be applied to GHC Core without modifications.
	We will elaborate on how we adapted these rules for Core in the second half of this section.
\item
	The type equality judgment $T_1 = T_2$ is used to unify two types $T_1$ and $T_2$.
\item
	The Sharing judgment $\sharerel{T}{\{T_1, T_2\}}$ enforces that all types have the same structure, but may have different type annotations within certain limitations.
	Most importantly, we enforce that the potential in the left hand side is greater or equal the sum of the potentials in the right hand side;
	In other words, the potential of $T$ is ``shared'' between $T_1$ and $T_2$.
	
	This property is used to ensure that potential can be redeemed only once.
	For example, consider a typing judgment \mbox{$\Gamma,\, x\of X \rtyps pq e : T$}, where the variable $x$ appears in expression $e$ multiple times.
	In a non-affine system, this would mean that any potential in $X$ could be redeemed once for each occurrence.
	But in our system, we instead use Sharing to split up the potential of $X$ between several subtypes $X_i$ which are then assigned to the various instances of $x$ in $e$.
	
	Note that the Sharing relation is defined for any number of types on the right hand side;
	In fact, the subtype relation $T_1 <: T_2$ is defined as a shorthand for $\sharerel{T_1}{\{T_2\}}$, a Sharing relation with only one type on the right hand side.
	
	As the Sharing judgment is a relation between different types and does not depend on the language syntax at all, we reuse the inference rules provided for JVFH \cite{TypeBasedCostAnalysisForLazyFunctionalLanguages} without any modifications.
\item
	The judgment $\lowertyps{T_1}{T_2}$ is another relation between different types and therefore can also be reused in our implementation without modifications.
	Just like Sharing, this judgment forces both $T_1$ and $T_2$ to have the same structure while giving some leeway to the type annotations.
	In particular, if $T_1$ and $T_2$ are some algebraic type $\mu X.\; \{\cdots\}$, then all recursive references $\thunkq qX$ within $T_1$ may have reduced thunk costs -- even zero -- compared to their counterparts in $T_2$.
	This is used as an optimization when analyzing recursive variable definitions, where we can assume that these thunks have already been evaluated.
	However, as we will discuss shortly, there were concerns about the correctness of this assumption and the operator is currently not used in our implementation.
\item
	The final kind of judgment in our system is linear constraints such as $v_1 \cdot c_1 + ... + v_n \cdot c_n \leq c_0$, where $v_i$ are numerical variables and $c_i$ are constants.
	We collect every linear constraint that comes up during the analysis into a single linear program, which is then passed to the LP solver.
\end{itemize}

As illustrated in this list, only the syntax driven type rules actually require modifications for GHC Core.
In the remainder of this section, we will list our adapted type rules and explain why and how they were modified.
For the original type rules, please refer to the original paper on JVFH \cite{TypeBasedCostAnalysisForLazyFunctionalLanguages} and the bachelor's thesis that introduced the type rule for mutually recursive definitions \cite{MutualRecursiveDefinitionsForATypeBasedCostAnalysisForLazyFunctionalLanguages}.

\begin{figure}
	\input{gc-tr-var}
	\bigskip
	\input{gc-tr-abs}
	\bigskip
	\input{gc-tr-appvar}
	\bigskip
	\input{gc-tr-letrec}
	\caption{Unmodified syntax driven type rules}
	\label{gc-unmodrules}
\end{figure}

Even among the syntax driven type rules, several rules could be reused with minor or no modifications, due to overlaps between the JVFH and GHC Core languages.
As depicted in Figure \ref{gc-unmodrules}, these encompass variables, abstraction, applying functions to variables, and recursive variable bindings.
Functionally, these four rules are identical to their respective counterparts in the JVFH systems.
However, the rules were renamed from \rnty{Var}, \rnty{Abs}, \rnty{App} and \rnty{LetAnd$\dagger$} to \ghctr{Var}, \ghctr{Abs}, \ghctr{AppVar} and \ghctr{LetRec};
And the latter has been adjusted for the minor syntactical differences between the JVFH $\lletand$ and the Core $\lletrec$ expressions.

However, \ghctr{LetRec} is also noteworthy as it is the only inference rule in our adapted type system that makes use of the $\lhd$ operator.
There exists a soundness proof of this rule \cite{MutualRecursiveDefinitionsForATypeBasedCostAnalysisForLazyFunctionalLanguages}, indicating that the usage of this operator should be correct;
But during our tests, we encountered a counter-example where this rule does allow for under-approximation:

\smallskip

\begin{minipage}{\textwidth}
	
	\hspace{2em}$\lletrec~ \mathit{zipWith} = \cdots ~\lin~ \llet~ \mathit{one} = \cdots ~\lin~ \llet~ \mathit{zero} = \cdots ~\lin$
	
	\hspace{2em}$\lletrec~ \mathit{fibs} = \lkey{Cons}~ zero~ \mathit{fibs}_1 ~;$
	
	\hspace{2em}\phantom{$\lletrec~$}$\mathit{fibs}_1 = \lkey{Cons}~ one~ \mathit{fibs}_2 ~;$
	
	\hspace{2em}\phantom{$\lletrec~$}$\mathit{fibs}_2 = \mathit{zipWith}~ \lkey{(+)}_{\lkey{Int}}~ \mathit{fibs}~ \mathit{fibs}_1$
	
	\hspace{2em}$\lin~ \mathit{fibs}$

\end{minipage}

\smallskip

For this (abridged) expression, which evaluates to an infinite list of Fibonacci numbers, our analysis was able to derive a list type with constant costs.
This is due to the $\lhd$ operator in the $\ghctr{LetRec}$ type rule, which allows us to treat the reference to $\mathit{fibs}_2$ in the definition of $\mathit{fibs}_1$ as if it had constant cost.
However, this is clearly incorrect -- $\mathit{fibs}_2$ is an infinite list which can never be fully evaluated, so the cost of the list has to be linear at least.
To work around this issue, we have disabled the $\lhd$ operator for now and simply use type unification in its place.
With this modification, our implementation will fail to analyze the code above, as the generated linear program is now unsolvable.
However, conceding failure is preferable to returning incorrect results.

\begin{figure}
	\input{gc-tr-app}
	\bigskip
	\input{gc-tr-cons}
	\bigskip
	\input{gc-tr-let}
	\bigskip
	\input{gc-tr-lit}
	\caption{Modified type rules}
	\label{gc-adaptrules}
\end{figure}

In Figure \ref{gc-adaptrules}, we list several rules which required larger modifications from the JVFH type system:

The application rule \ghctr{App} is a generalization of the original \rnty{App} rule.
In JVFH, the arguments of function applications always have to be variables, but in GHC Core, this is not necessarily the case.
Therefore, we adapted this rule to allow for any expression in the argument.
However, if the argument actually is a variable, using this generalized rule can artificially increase the thunk cost of the variable.
In those cases, we instead use the previously mentioned \ghctr{AppVar} rule, which is an unmodified copy of the original \rnty{App} rule, to prevent this undesired side-effect.

\ghctr{Cons} can be considered a replacement of the original \rnty{LetCons} and \rnty{Cons} rules.
JVFH handles structure allocation via a $\lletcons$ expression, which fulfills multiple purposes at once:
First, it fully applies the constructor to all of its arguments;
This is simple, because all arguments have to be variables, so this can be done by looking up their types in the context.
In GHC Core, the arguments can be more complex, so we defer handling the arguments to the \ghctr{App} and \ghctr{AppVar} rules instead.
Second, the resulting structure is immediately bound to a variable in JVFH.
However, in GHC Core, a structure can sometimes be used immediately without binding it to a variable first.
This is generally the case when the constructor does not have any fields, such as the ``empty list'' constructor of a list type.
Because of these differences, we chose to process constructors individually, instead of aggregating them with surrounding application and binding expressions.

The \ghctr{Let} rule is a simplified version of the original \rnty{Let$\dagger$} rule.
Unlike JVFH, the $\llet$ expression in GHC Core does not support recursive definitions -- the $\lletrec$ expression has to be used instead.
Our rule was adapted to reflect this difference.

\ghctr{Lit} is an entirely new rule introduced to type literals.
As we do not distinguish between different primitive types such as \texttt{Int\#}, \texttt{Float\#} or \texttt{Char\#}, we simply type these as the empty algebraic type $\mu X. \{\}$.

\begin{figure}
	\input{gc-tr-casealg}
	\bigskip
	\input{gc-tr-caselit}
	\caption{Adapted type rules for {\lcase} expressions}
	\label{gc-caserules}
\end{figure}

The rules in Figure \ref{gc-caserules} cover two different kinds of pattern matches, namely on algebraic types and on primitive types:

\ghctr{CaseAlg} was extended from the original \rnty{Match} rule to account for some additional features of Core's $\lcase$ expression which are not available to JVFH's $\lmatch$;
Namely, binding the value of the scrutinee to a variable, and an additional \lkey{default} case which is used when none of the other cases match.
\ghctr{CaseLit}, then, was heavily simplified from our own \ghctr{CaseAlg} rule, as primitive types cannot have fields and as we do not distinguish between different primitive values.

\begin{figure}
	\input{gc-tr-tyabs}
	\bigskip
	\input{gc-tr-tyapp}
	\bigskip
	\input{gc-tr-tylet}
	\caption{New type rules for handling type expressions}
	\label{gc-tyrules}
\end{figure}

Finally, we introduced the rules listed in Figure \ref{gc-tyrules}, which offer very limited support for type abstraction and type application.
Unfortunately, as mentioned earlier, our type system currently does not support polymorphism yet, as this feature is also missing in JVFH, and extending the type system with major new features was not within the scope of our work.
Instead, we opted to simply ``ignore'' type abstraction and application for now.
To see why this can be convenient, consider the \texttt{repeat} example discussed earlier:

\smallskip

\begin{minipage}{\textwidth}
	
	\hspace{2em}\verb$repeat x = let xs = x : xs in xs$
	
	\hspace{2em}\verb$it = repeat 1 :: [Int]$

\end{minipage}

\smallskip

We do not explicitly specify a type for the function \texttt{repeat};
%Therefore, GHC will automatically infer the polymorphic type \mbox{``$\forall a.~ a \to [a]$''} and introduce the appropriate type abstractions and applications into the generated Core code.
Therefore, GHC will automatically infer the polymorphic type \mbox{``\texttt{forall a.\ a -> [a]}''} and introduce the appropriate type abstractions and applications into the generated Core code.
Without our type rules from Figure \ref{gc-tyrules}, our analysis would fail when it encounters any of these expressions.
%However, if we instead simply ignore them, we will then unify the type variable $a$ with $\lkey{Int}$, which essentially turns the function into a monomorphic function of type \mbox{``$\lkey{Int} \to [\lkey{Int}]$''.}
However, if we instead simply ignore them, we will then unify the type variable \texttt{a} with \texttt{Int}, which essentially turns the function into a monomorphic function of type \mbox{``\texttt{Int -> [Int]}''.}
An additional wrapper $\typAr{\mu X. \{\}}0{C}$ around the actual type $C$ is introduced as an annotation to inform the user about any ignored type abstractions.

However, this should be considered a workaround, and will only work if every polymorphic function is used monomorphically.
If it is used polymorphically, this will lead to the unification of two incompatible types;
And if it is never used at all, the resulting type will contain free type variables, which is not valid in our system.

\section{Implementation}

\begin{figure}
	\centering
	\begin{tikzpicture}
		\begin{scope}[every node/.style={rectangle, draw=black, fill=white}]
			\node[draw=none] (Start) at (0,0) {.hs file};
			\node[right = 1cm of Start] (GHC) {GHC compiler};
			\node[below = 5mm of GHC] (Plugin) {GHC plugin};
			\node[draw=none, right = 1cm of Plugin, align=center] (Core) {Core\\program};
			\node[draw=none, right = 1cm of Core, align=center] (Constr) {initial typing\\judgment};
			\node[right = 1cm of Constr, align=center] (Solver) {type system};
			\node[draw=none, above = 1cm of Solver] (Cts) {judgments};
			\node[draw=none, inner sep = 0, below = 1cm of Solver] (Result) {};
			\node[draw=none, left = 5mm of Result, align=center] (Tp1) {typing with\\variable annotations};
			\node[draw=none, right = 1cm of Result] (LP) {LP};
			\node[below = 1cm of LP] (LPS) {LP solver};
			\node[draw=none, below = 1cm of Tp1, align=center] (Tp2) {typing with\\value annotations};
		\end{scope}
		
		\begin{scope}[-{Latex[length=3mm,width=2mm]}]
			\draw (Start) -> (GHC);
			\draw (Plugin) -> (Core);
			\draw (Core) -> (Constr);
			\draw (Constr) -> (Solver);
			\draw (Solver) to[bend left = 40] (Cts);
			\draw (Cts) to[bend left = 40] (Solver);
			\draw (Solver) -> (Tp1);
			\draw (Solver) -> (LP);
			\draw (LP) -> (LPS);
			\draw (LPS) -> (Tp2);
			\draw (Tp1) -> (Tp2);
		\end{scope}
		
		\begin{scope}[-{Latex[length=1.5mm,width=1mm]}]
			\draw (GHC) -> (Plugin);
			\draw (Plugin) -> (GHC);
		\end{scope}
	\end{tikzpicture}
	\caption{Outline of the architecture of our analysis}
	\label{anlysisarchitecture}
\end{figure}
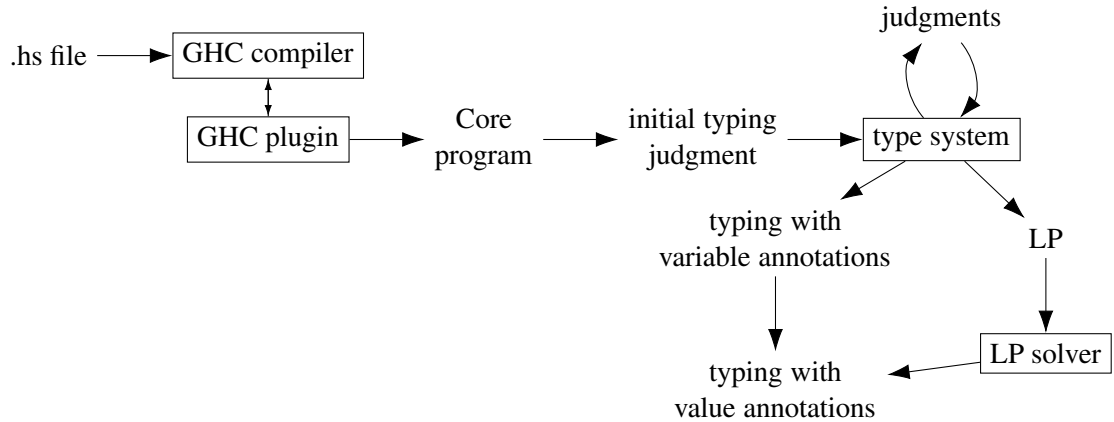

The architecture of our tool is shown in Figure \ref{anlysisarchitecture}.
We initiate the analysis by calling the GHC compiler on a given Haskell module.
After the code was compiled to GHC Core, it is then fetched by our GHC plugin and passed to the actual analysis.
Note that this Core ``program'' is actually a list of variable bindings;
However, we can convert this list into a single expression by simply nesting $\llet$ and $\lletrec$ expressions.
This allows us to generate an initial typing judgment, or ``constraint'', which needs to be proven.

A constraint solver is used to apply the appropriate type rules to each judgment, and to keep track of the judgments that already have been or still need to be proven.
Using a central ``control unit'' for managing derived judgments like this has several advantages;
For example, this allows us to print a trace when we encounter a typing error, or to output the full derivation tree after the analysis has completed.
This can be used as an evidence to support the correctness of the analysis result, or as a tool for debugging.

In most cases, there is no ambiguousness regarding which inference rule to apply on any given judgment;
However, structural type rules are an exception, as any of them may be applied to any typing judgment.
For our implementation, we therefore need to find a way to make these rule applications deterministic.
A naive solution for this problem is to apply every structural once to every generated typing judgment before applying the next syntax-driven type rule.
However, this needlessly increases the size of the derivation tree, as most of these rule applications would not add anything to the strength of our type system.

Instead we reused the same heuristics as described in section ``4.5 Experimental Results'' of the original paper on JVFH. \cite{TypeBasedCostAnalysisForLazyFunctionalLanguages}
For example, consider the \rnty{Prepay} rule, which allows us to prepay the cost of a variable in the context.
This is used to simulate the fact that thunks are evaluated at most once, even when they are referenced multiple times in an expression.
According to Jost et al., it suffices to apply this rule once whenever a new variable is inserted into the context of a typing judgment.
In our system, this applies to the \ghctr{Abs}, \ghctr{LetRec}, \ghctr{Let} and \ghctr{CaseAlg} rules.
Note that the \ghctr{CaseLit} also introduces a new variable, but its thunk cost is always zero, so prepaying would be redundant.

\begin{sloppypar}
For the most part, we found the given heuristics to be sufficient, with one exception:
In the \ghctr{CaseAlg} and \ghctr{CaseLit} rules, the number of resources remaining after evaluating the scrutinee is also used as the number of resources available before evaluating the respective cases.
However, for most scrutinees, our type system will set this number to a low value which generally does not suffice for evaluating the cases.
Therefore, we found it necessary to apply the \rnty{Relax} rule to the scrutinee judgment.
This allows us to artificially increase the number of resources available both before and after the evaluation by any value as needed.
The intuition is that these additional resources are not used during the evaluation of the scrutinee;
Therefore, they are available before its evaluation and also remain available afterwards.
\end{sloppypar}

In addition to the syntax-driven type rules mentioned in the previous section, we also had to hard-code the types of some specific variables.
Most notably, these include operations on primitive types such as \texttt{Int\#} or \texttt{Float\#}.
Usually, variables simply resolve to actual Haskell code, which we could analyze to derive an annotated type and an LP over its annotations.
However, primitive operators do not correspond to any Haskell code;
Instead, they are detected by the GHC compiler and translated into an appropriate opcode for the target machine of the compilation.
Since these operators are hard-coded into the compiler, we also need to hard-code them into our analysis.
We introduce an additional cost constant $\RK{prim}$ which is used for every execution of a primitive operation.

We also hard-coded several other variables from the Prelude, including dictionaries and operators on boxed primitive types such as \texttt{Int} and \texttt{Float}.
This is a temporary solution introduced only because our implementation currently does not support multi-module programs yet.
In the future, references to other modules -- including the Prelude -- should automatically be detected and handled correctly, rendering this workaround obsolete.

Finally, for solving the linear program we make use of glpk-hs, which previously was also used in the implementation of the JVFH type system.
This package provides Haskell bindings for the GNU Linear Programming Kit.
However, we have introduced an additional abstraction layer, so the use of this library is transparent to the majority of our code.
Thus, it should be possible to replace the LP solver without much effort if required.

\section{Evaluation}
\label{evaluation}

Our implementation can analyze a substantial subset of Haskell.
For example, consider the following simple Haskell module, which contains examples of several concepts that are common to Haskell, but nonexistent in JVFH, such as list comprehension and function applications with non-variable arguments:

\smallskip

\begin{minipage}{\textwidth}
	
	\hspace{2em}\verb$module Example where$
	
	\hspace{2em}\verb$import Prelude hiding (map, repeat)$
	
	\hspace{2em}\verb$repeat x = xs where xs = x : xs$
	
	\hspace{2em}\verb$map f xs = [f x | x <- xs]$
	
	\hspace{2em}\verb|it = map (+1) $ repeat 1 :: [Int]|
	
\end{minipage}

\smallskip

When applied to this module, our analysis will successfully generate the following typing for the variable \texttt{it}:

\hspace{2em}$\rtyps90 \mathit{it} : \mu X.~ \{\lkey{[]} : (0, []) ~|~ \lkey{(:)} : (0, [\thunkq1{\lkey{Int}}, \thunkq3X])\}$

This means that evaluating this expression to weak head normal form will induce a onetime cost of 9 allocations at most, and up to 3 more allocations for each list node accessed.
Furthermore, accessing any of the list elements will also evoke one additional allocation.

For comparison, consider the following translation of the previous code to JVFH:

\smallskip

\begin{minipage}{\textwidth}
	
	\hspace{2em}\verb$let repeat = \x -> letcons xs = Cons(x,xs) in xs$
	
	\hspace{2em}\verb$in let map = \f xt -> match xt with$
	
	\hspace{2em}\verb$           Nil() -> letcons r = Nil() in r$
	
	\hspace{2em}\verb$         | Cons(x,xs) -> let y = f x$
	
	\hspace{2em}\verb$                         in let ys = map f xs$
	
	\hspace{2em}\verb$                            in letcons r = Cons(y,ys)$
	
	\hspace{2em}\verb$                               in r$
	
	\hspace{2em}\verb$   in let one = 1$
	
	\hspace{2em}\verb$      in let ones = repeat one$
	
	\hspace{2em}\verb$         in let inc = \x -> let r = one + x in r$
	
	\hspace{2em}\verb$            in let it = map inc ones$
	
	\hspace{2em}\verb$               in it$
	
\end{minipage}

\smallskip

From this expression, the original JVFH analysis will infer the following typing:

\hspace{2em}$\rtyps{10}0 \mathit{it} : \mu X.~ \{\lkey{[]} : (0, []) ~|~ \lkey{(:)} : (0, [\thunkq1{\lkey{Int}}, \thunkq3X])\}$

Note that the typing is similar, but not identical.
The specific values used as type annotations may differ, due to differences in how the code is represented in JVFH and GHC Core, respectively;
and the LP solver also has some freedom in how values are assigned to variables.

This result is typical of most comparisons between JVFH and Haskell analyses.
As we have not formally proven the correctness of our adapted system, we used JVFH analysis results as a basis for testing the viability of our own analysis.
For this purpose, we translated several JVFH example programs published as part of a JVFH online demo\footnote{The JVFH online demo is available at: \url{http://kashmir.dcc.fc.up.pt/cgi/lazy.cgi}} to Haskell.
We then ran our analysis on this code and compared our results to those from the online demo.
While we obtained similar results in most cases, there were a few noteworthy cases where our results were entirely different.

One of these is the following code which is a simplified version of the ``list fusion'' example from the online demo.
For brevity, we will only provide our Haskell translation of the code:

\smallskip

\begin{minipage}{\textwidth}
	
	\hspace{2em}\verb$map1 f [] = []$
	
	\hspace{2em}\verb$map1 f (x:xs) = f x : map1 f xs$
	
	\hspace{2em}\verb$map2 f [] = []$
	
	\hspace{2em}\verb$map2 f (x:xs) = f x : map2 f xs$
	
	\hspace{2em}\verb|lhs f g xs = map1 f $ map2 g xs|
	
\end{minipage}

\smallskip

Note that we provide duplicate definitions of the \texttt{map} function for each usage within the \texttt{lhs} function.
This is a workaround commonly used in the JVFH examples, which allows us to assign different types with different annotations to each usage of the function.
Using this trick, the original JVFH version of the code above can be analyzed without any issues.
However, when we attempt to analyze our Haskell translation, the GHC compiler detects this duplication and removes it, replacing all references to \texttt{map2} with \texttt{map1}.
As a result, our analysis of the \texttt{lhs} function will generate an unsolvable linear program and then fail.
This is beyond our control, as we can only edit the Haskell code, and the translation to GHC Core is left entirely to the compiler.
%However, note that this duplicate code elimination seems to be very fragile;
%For example, when we simply replaced our implementation of both \texttt{map} function with a list comprehension (as in the example before), GHC will keep the duplication intact during the compilation and our analysis succeeds.
%TODO: ist das sonderlich relevant/interessant?

However, this property also facilitated the discovery and elimination of an issue with a type rule that had previously been considered sound.
We found this issue while analyzing the following definition of the Fibonacci sequence:

\smallskip

\begin{minipage}{\textwidth}
	
	\hspace{2em}\verb$zipWith f (x:xs) (y:ys) = f x y : zipWith f xs ys$
	
	\hspace{2em}\verb$fibs = 0 : 1 : zipWith (+) fibs fibs' where (_:fibs') = fibs$
	
\end{minipage}

\smallskip

Initially, our analysis erroneously was able to derive a constant cost for this sequence, meaning that it would be possible to fully evaluate the entire infinite list.
We have previously detailed the cause and our fix of this issue in our discussion of the \ghctr{LetRec} type rule in section \ref{typerules}.
However, this issue was exposed only because GHC translated our Haskell code to Core using $\lletrec$ for mutual recursion;
The original JVFH version of this code did not contain any $\lletand$ expressions -- which is the JVFH equivalent of $\lletrec$ -- and therefore the erroneous type rule was not triggered during analysis.
However, we were able to reproduce this error in the JVFH analysis by manually translating the Core code to JVFH, proving that this is an issue with the original rule, and not with our own implementation.
While our fix causes the analysis of this Fibonacci code to fail due to an unsolvable linear program, this is preferable over returning an incorrect analysis result.

\section{Outlook}

The tool described in this paper was implemented as part of a master's thesis. \cite{ImplementationOfAnAutomatedAmortizedAnalysisOnGHCCoreAsCompilerPlugin}
The goal of this thesis was to provide an initial implementation, which can then be used as a basis for future improvements.
While our analysis already works on a substantial subset of Haskell, there still are several flaws that severely inhibit its usefulness.
In this section, we will discuss some of the shortcomings that may be used as a starting point for any future work.

The biggest issue of our current work is the lack of support for polymorphism.
As we previously discussed in section \ref{ghcarchitecture}, we specifically chose GHC Core over STG for our analysis with future support for polymorphism in mind;
But as our type system is based on JVFH, which does not feature polymorphism at all, this feature is also missing in our current implementation.
While the type syntax does include type variables, these may only be used as recursive references in an algebraic type;
And several inference rules in the JVFH type system (which we currently reuse without alterations) were written with this assumption in mind.
Therefore, providing full support for polymorphism may require large modifications to our type system.

We currently also do not support the ``Cast'' and ``Coercion'' expressions of GHC Core.
These are used by the compiler to support types defined via \texttt{newtype}, as well as several Haskell language extensions such as GADT, associated types, and functional dependencies. \cite{SystemFWithTypeEqualityCoercions}
Therefore, none of these features are currently supported in our analysis.
As of yet, we have not investigated which modifications to our type system would be necessary to support these two expressions.

Multi-module programs also are not supported yet;
Therefore, we cannot analyze any expressions that contain variables imported from different modules, including the \texttt{Prelude}.
In some cases, however, the compiler may automatically inline function calls, circumventing this limitation.
This is the reason why we can use the operators \texttt{+} and \texttt{\$} in the example codes in the previous section.

We also have not formally proven the soundness of our adapted type system.
For this purpose, it would be necessary to define the cost-annotated operational semantics for GHC Core;
and to show that type annotations calculated by our analysis are always greater or equal to the costs determined by the operational semantics.
As we have previously discussed in sections \ref{typerules} and \ref{evaluation}, it may also be necessary to revisit the soundness proof for the original \rnty{LetAnd$\dagger$} rule \cite{MutualRecursiveDefinitionsForATypeBasedCostAnalysisForLazyFunctionalLanguages}.

\section{Conclusion}

%Introduction
In this paper, we have presented an automated amortized analysis on the resource usage of Haskell programs.
% Related Work
Our work is based on a previous paper by Jost et al.\ \cite{TypeBasedCostAnalysisForLazyFunctionalLanguages}, which proposed a type-based analysis on an artificial language called JVFH.
%Similar previous attempts were generally focused on languages with strict evaluation \cite{TODO}, or on artificial languages which are not in common use \cite{TODO}.
% GHC Architecture
We use the plugin API of the GHC compiler to translate any given Haskell code into a heavily simplified representation called GHC Core.
% Language & Type Syntax
Due to the large similarities to JVFH, we can reuse most of the definitions and inference rules from the JVFH system.
% Type Rules
Only a subset of the type rules were modified to take minor syntactic and semantic differences between the two languages into account.
% Implementation
For the most part, our implementation will return similar results to those produced by the original JVFH analysis; % Evaluation
% Outlook
However, there still are several shortcomings, which make our current implementation impractical for everyday use.
Most notably, certain features such as polymorphism and \texttt{newtype} are not supported yet and were deferred to future work.
% Conclusion

\bibliographystyle{eptcs}
\bibliography{paper}
\end{document}